\documentclass[letterpaper,twocolumn,10pt]{ilass}
\def\thepage{}

\usepackage{amssymb}
\usepackage{amsmath}

\usepackage{ifpdf}

\ifpdf
\usepackage[pdftex]{graphicx}
\else
\usepackage{graphicx}
\fi

\usepackage{subfigure}
\usepackage{graphicx}
\usepackage{textcomp}
\usepackage[nameinlink, capitalize]{cleveref}
\usepackage[export]{adjustbox}

\allowdisplaybreaks

\graphicspath{{./figures/}}

\title{
    \vspace{-0.25in} 
    {\small \em ILASS-Americas 29th Annual Conference on Liquid Atomization and Spray Systems, Atlanta, GA, May 2017} 
    \newline \newline
    \large {\bf Modeling and Simulation of Diesel Injection at Transcritical Conditions}
}
        
\author{
    \large
    Peter C. Ma, Matthias Ihme\footnote{Corresponding Author: mihme@stanford.edu}\\
    Department of Mechanical Engineering\\
    Stanford University\\
    Stanford, CA 94305\\
    \\
    Luis Bravo\\
    Multiphase Combustion Computational Research\\
    Propulsion Division, Vehicle Technology Directorate\\
    U.S. Army Research Laboratory\\
    Aberdeen Proving Ground, MD 21005    
}

\date{\normalsize  \centerline{\bf Abstract} \vspace{0.05in}
\begin{minipage}{6.5in}
\normalsize
The need for improved engine efficiencies has motivated the development of high-pressure combustion systems, in which operating conditions achieve and exceed critical conditions. Associated with these conditions are large thermodynamic gradients and strong variations in transport properties as the
fluid undergoes mixing and phase transition. Accurately simulating these real-fluid environments remains a main challenge. Different modeling approaches have been employed, which can be categorized as diffused and sharp interface methods. The objective of this study is to examine the diffused interface method for simulating diesel-fuel injection at conditions related to the supercritical regime. To this end, a recently developed compressible real-fluid solver for transcritical conditions is employed. Simulations of an ECN-relevant diesel-fuel injector are performed and predictions for instantaneous and statistical flow-field results are compared against available measurements. It is expected that results from this analysis will be useful in identifying limitations of current modeling techniques and in improving physical and numerical models for high-pressure injection systems.
\end{minipage} \vspace{-0.25in}
}

\begin{document}

\ifpdf
\DeclareGraphicsExtensions{.pdf, .jpg}
\else
\DeclareGraphicsExtensions{.eps, .jpg}
\fi

\maketitle

\clearpage
\pagenumbering{arabic}
\setcounter{page}{2}
\section{Introduction} 
\begin{figure}[!htb!]
    \centering
    \includegraphics[width=0.95\columnwidth]{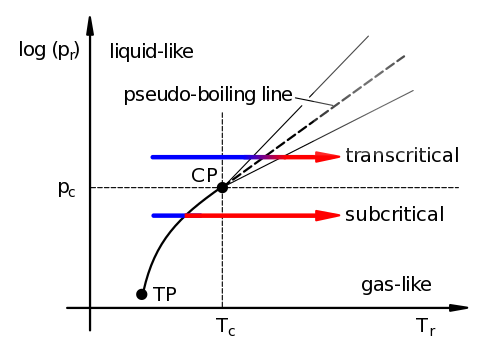}
    \caption{Illustration of subcritical and transcritical injection processes in a $p_r$-$T_r$ diagram~\cite{banuti2017seven}. CP denotes the critical point, and TP refers to the triple point.\label{fig:subsupercritical}} 
\end{figure}

The thermodynamic analysis of diesel engines shows that a higher chamber pressure will yield not only higher energy output but also higher engine efficiency. The need for improved engine efficiencies has motivated substantial research interests in high-pressure combustion systems, in which operating conditions achieve and exceed critical conditions. In diesel engines, typically a subcritical liquid fuel is injected into a supercritical ambient gaseous environment, and the fuel jet goes breakup, heating, and mixing processes before combustion. This process is referred to as a transcritical injection, which is schematically shown in \cref{fig:subsupercritical} in comparison with a classical subcritical injection process. At elevated pressure, the fluid properties exhibit liquid-like densities and gas-like diffusivities, and the surface tension and enthalpy of vaporization approach zero \cite{yang2000modeling}. This phenomenon has been shown by recent experimental works~\cite{mayer2000injection,oschwald2006injection,chehroudi2012recent} and theoretical analysis~\cite{dahms2013transition,dahms2016understanding}. However, these complex processes are still not fully understood. Therefore, to provide insights into the high-pressure combustion systems, accurate and robust simulation tools are required for the characterization of supercritical and transcritical flows where large density gradients and thermodynamic anomalies occur.

The transcritical nature of flows in a diesel injection process has motivated several recent studies to utilize the diffused interface method for the modeling of the injection sequence. In contrast to a sharp interface method, such as a volume of fluid method, where interfaces are explicitly tracked or resolved in the computation domain, the diffused interface method artificially diffuses the interfaces. This is particularly attractive for transcritical flows where interfaces are not present. However, it remains an open research question whether interfacial flows or droplets exist under conditions relevant to real applications~\cite{yang2000modeling,banuti2017seven}. Lacaze~et~al.~\cite{lacaze2015analysis} performed a large-eddy simulation (LES) using a real-fluid state equation to analyze the details of the transient mixing and the processes leading to autoignition during diesel injection. It was found that the large density ratio leads to significant penetration of the jet and enhanced entrainment effects. Ma~et~al.~\cite{ma2014supercritical} developed a numerical scheme for the robust simulation under diesel relevant conditions. Knudsen~et~al.~\cite{knudsen2016compressible} employed a compressible Eulerian numerical model to describe the liquid fuel injection process under consideration of the internal nozzle flow and compressibility effects. Matheis~and~Hickel~\cite{matheis2016multi} developed a thermodynamic model where phase separation is considered through vapor liquid equilibrium calculations to enable the robustness of the diffused interface method when a fully conservative scheme is used. 

Several challenges require considerations for the modeling of transcritical flows using the diffused interface method. One is the large density gradients caused by the liquid-like density in the dense fluid region and the gas-like density in the supercritical ambient gas region. A typical treatment is to introduce numerical dissipation locally where large density gradients occur~\cite{selle2010large,ruiz2012unsteady,terashima2012approach,hickey2013large,ma2017entropy}. Another difficulty arises from spurious pressure oscillations that are generated from nonlinearities of the real-fluid equation of state. These oscillations arise when fully conservative schemes are applied to multicomponent flows~\cite{abgrall2001computations}. Schmitt~et~al.~\cite{ruiz2012unsteady,schmitt2010large} derived a quasi-conservative scheme by relating the artificial dissipation in the conservation equations and setting the pressure differential to zero. Terashima~et~al.~\cite{terashima2012approach,kawai2015robust} solved the pressure equation instead of the energy equation. Ma~et~al.~\cite{ma2017entropy} developed a double-flux model with entropy-stability to eliminate spurious pressure oscillations in transcritical flows.

In this study, a recently developed compressible real-fluid solver~\cite{ma2014supercritical,ma2017entropy} is used for the simulation of a diesel injection process under realistic engine conditions. The governing equations of the conservation laws will be presented followed by the thermodynamic and transport models used for the description of the transcritical flows. The numerical methods employed in this study will be briefly discussed. These simulation results will be compared to available experimental measurements to examine the performance of the current numerical scheme and the diffused interface method for the prediction of the flow structures and mixing behaviors in a diesel injection process. The paper finishes with conclusions.

\section{Mathematical Model}
\subsection{Governing Equations}
The governing equations are the conservation laws for mass, momentum, total energy, and species, taking the following form:
\begin{subequations}
    \label{eqn:governingEqn}
    \begin{align}
        \frac{\partial \rho}{\partial t} +\nabla \cdot (\rho \boldsymbol{u}) &= 0\,,\\
        \frac{\partial (\rho \boldsymbol{u})}{\partial t} + \nabla \cdot (\rho \boldsymbol{u} \boldsymbol{u} + p \boldsymbol{I}) &= \nabla \cdot \boldsymbol{\tau} \,,\\
        \frac{\partial  (\rho e_t)}{\partial t} + \nabla \cdot (\boldsymbol{u}(\rho e_t + p)) &= \nabla \cdot (\boldsymbol{\tau} \cdot \boldsymbol{u}) -\nabla \cdot \boldsymbol{q} \,,\\
        \frac{\partial (\rho Y_k)}{\partial t} + \nabla \cdot (\rho \boldsymbol{u} Y_k) &= \nabla \cdot (\rho D_k \nabla Y_k) \,,
    \end{align}
\end{subequations}
where $\rho$ is the density, $\boldsymbol{u}$ is the velocity vector, $p$ is the pressure, $e_t$ is the specific total energy, $Y_k$ is the mass fraction of species $k$, $D_k$ is the diffusion coefficient for species $k$, and $k = 1\,, \dots, N_S -1$ in which $N_S$ is the number of species. The viscous stress tensor and heat flux are written as:
\begin{subequations}
    \begin{align}
        \boldsymbol{\tau} &=  \mu \left[ \nabla \boldsymbol{u} + (\nabla \boldsymbol{u})^T \right] -\frac{2}{3}\mu (\nabla \cdot \boldsymbol{u}) \boldsymbol{I} \;,\\
        \boldsymbol{q} &= - \lambda \nabla T - \rho \sum_{k = 1}^{N_S} {h}_k D_k \nabla Y_k\;,
    \end{align}
\end{subequations}
where $T$ is the temperature, $\mu$ is the dynamic viscosity, $\lambda$ is the thermal conductivity, and ${h}_k$ is the partial enthalpy of species $k$. The total energy is related to the internal energy and the kinetic energy:
\begin{equation}
    \rho e_t = \rho e + \frac{1}{2}\rho \boldsymbol{u} \cdot \boldsymbol{u}\,.
\end{equation}
The system is closed with a state equation, which is here written in the general form as,
\begin{equation}
    p = f(\rho, e, Y_k)\,.
\end{equation}
The formulation of the state equation are discussed next.

\subsection{Equation of State}
For computational efficiency and the accurate representation near the critical point \cite{miller2001direct}, the Peng-Robinson cubic state equation~\cite{peng1976new, poling2001properties} is used in this study, which can be written as:
\begin{equation}
    \label{EQ_PR_EQUATION_P}
    p = \frac{R T}{v - b} - \frac{a}{v^2+2bv-b^2}\,,
\end{equation}
where $R$ is the gas constant, $v$ is the specific volume, and the coefficients $a$ and $b$ are dependent on temperature and composition to account for effects of intermolecular forces.

The extended corresponding-states principle and pure fluid assumption for mixtures are adopted in this study~\cite{ely1981prediction, ely1983prediction}. Mixing rules are applied for coefficients $a$ and $b$ in \cref{EQ_PR_EQUATION_P}:
\begin{subequations}
    \label{EQ_PR_EQUATION_ab}
    \begin{align}
        a &=  \sum^{N_S}_{i=1}\sum^{N_S}_{j=1} X_i X_j a_{ij}\,,\\
        b &= \sum^{N_S}_{i=1} X_i b_i\,,
    \end{align}
\end{subequations}
where $X_i$ is the mole fraction of species $i$. The coefficients $a_{ij}$ and $b_i$ are evaluated using the recommended mixing rules by Harstad et al.~\cite{harstad1997efficient}:
\begin{subequations}
    \label{EQ_PR_EQUATION_alphabeta}
    \begin{align}
        a_{ij} &= 0.457236\frac{(R T_{c,ij})^2}{p_{c,ij}} \left[1+c_{ij}\left(1-\sqrt{\frac{T}{T_{c,ij}}}\right)\right]^2\;,\\
        b_{i} &= 0.077796 \frac{R T_{c,i}}{p_{c,i}}\;,\\
        c_{ij} &= 0.37464 + 1.54226\omega_{ij} - 0.26992\omega^2_{ij}\;.
    \end{align}
\end{subequations}
The critical mixture temperature, pressure, molar volume, compressibility, and acentric factor are determined according to Harstad et al.~\cite{harstad1997efficient}.

Procedures for evaluating thermodynamic quantities such as internal energy, specific heat capacity and partial enthalpy using the Peng-Robinson state equation are described in detail in Hickey~et~al.~\cite{hickey2013large} and Ma~et~al.~\cite{ma2017entropy,ma2017numerical}.
\subsection{Transport Properties}\label{sec:transport}
The dynamic viscosity and thermal conductivity are evaluated using Chung's method with high-pressure correction \cite{chung1984applications, chung1988generalized}. However, the evaluation of the viscosity in this model in known to produce oscillations for certain mixtures~\cite{hickey2013large,ruiz2015numerical}. To resolve this issue, in the present study the acentric factor is set to zero when evaluating the viscosity to avoid the anomalies.

Takahashi's high-pressure correction~\cite{takahashi1975preparation} is used to evaluate binary diffusion coefficients. Since only binary mixtures are considered in this work, the binary diffusion coefficient is the only property needed for the species equations and the evaluation of diffusion coefficients is exact. For more complex mixtures, a mixture-averaged or multi-component treatment of the diffusion coefficient can be adopted~\cite{kee2005chemically}.
\subsection{Numerical Schemes}
A finite volume approach is utilized for the discretization of the system of equations, \cref{eqn:governingEqn}:
\begin{equation}
    \frac{\partial U}{\partial t} V_{cv} + \sum_f F^e A_f = \sum_f F^v A_f\,,
\end{equation}
where $U = [\rho, \rho u, \rho v, \rho w, \rho e_t, \rho Y_k]^T$ is the vector of conserved variables, $F^e$ is the face-normal Euler flux vector, $F^v$ is the face-normal viscous flux vector which corresponds to the right-hand side (RHS) of \cref{eqn:governingEqn}, $V_{cv}$ is the volume of the control volume, and $A_f$ is the face area. A strong stability preserving 3rd-order Runge-Kutta (SSP-RK3) scheme \cite{gottlieb2001strong} is used for time integration. A Strang-splitting scheme~\cite{strang1968construction} is applied in this study to separate the convection operator from the remaining operators of the system. For LES calculations, the governing equations are Favre-filtered and a Vreman subgrid-scale model~\cite{vreman2004eddy} is used as turbulence closure.

The convective flux is discretized using a sensor-based hybrid scheme~\cite{khalighi2011unstructured} in which a high-order, non-dissipative scheme is combined with a low-order, dissipative scheme to minimize the numerical dissipation introduced. A central scheme which is fourth-order on uniform meshes is used along with a second-order ENO scheme for the hybrid scheme. A density sensor~\cite{ma2014supercritical, ma2017entropy} is adopted in this study. Due to the large density gradients present under transcritical conditions, an entropy-stable flux correction technique~\cite{ma2017entropy} is used to ensure the physical realizability of the numerical solutions and to dampen  non-linear instabilities in the numerical schemes. 

Due to the strong non-linearity inherent in the real-fluid state equations, spurious pressure oscillations will be generated when a fully conservative scheme is used~\cite{terashima2012approach, ma2017entropy}, and severe oscillations in the pressure field could possibly lead the solver to diverge. A double-flux method~\cite{ma2014supercritical,ma2017entropy,ma2016numerical} is extended to the transcritical regime to eliminate the spurious pressure oscillations. The effective specific heat ratio based on the speed of sound is frozen both spatially and temporally for a given cell when the fluxes at the faces are evaluated. The conservation error in total energy was shown to converge to zero with increasing resolution~\cite{ma2017entropy,ma2016numerical}.
\section{Results and Discussion}
In this study, the Spray A configuration~\cite{pickett2011engine} is considered, representing a benchmark target of the Engine Combustion Network. The single hole diesel injection is operated with pure n-dodecane at a rail pressure of 150~MPa. Liquid n-dodecane fuel is injected at 363~K through a nozzle with a diameter of 0.09~mm into a 900~K ambient environment at a pressure of 6.0~MPa. The non-reacting case is considered and the ambient gas consists of pure nitrogen. At these conditions, the liquid n-dodecane undergoes a transcritical injection process as illustrated in~\cref{fig:subsupercritical}, where the liquid fuel is heated and mixes with the ambient gaseous environment.

A cylindrical computational domain is used in this study with a diameter of 20~mm and a length of 80~mm. The injector geometry is not included in the computation domain. The mesh is clustered in the region  near the injector along the shear layers, and stretched in downstream and radial directions. The minimum grid spacing is 4~\textmu m, using approximately 20 grid points across the injector nozzle. The total cell count of the mesh is 4.6 million.

Fuel mass flux and temperature are prescribed at the injector nozzle using the time-dependent rate of injection modeled using the CMT virtual injection rate generator~\cite{pickett2011engine}, recommended by the ECN with default input parameters for the Spray A case (150~MPa, 6~MPa, 0.0894~mm, 0.90, 713.13~kg/m$^3$, and 1.50~ms for injection pressure, back pressure, outlet diameter, discharge coefficient, fuel density and injection time, respectively). Plug flow velocity profile is applied at the nozzle exit without synthetic turbulence. The pressure is prescribed at 6~MPa at the outlet. Adiabatic boundary conditions are applied at all walls. The numerical simulation is initialized with pure nitrogen at 900~K with zero velocity and a pressure of 6~MPa. A CFL number of 1.0 is used during the simulation and a typical time step is about 0.6~ns. The injection process is simulated over a duration of 1.2~ms.

\begin{figure*}[!htb!]
    \centering
    \subfigure{
        \includegraphics[width=0.45\textwidth,trim={75 32 85 15},clip=]{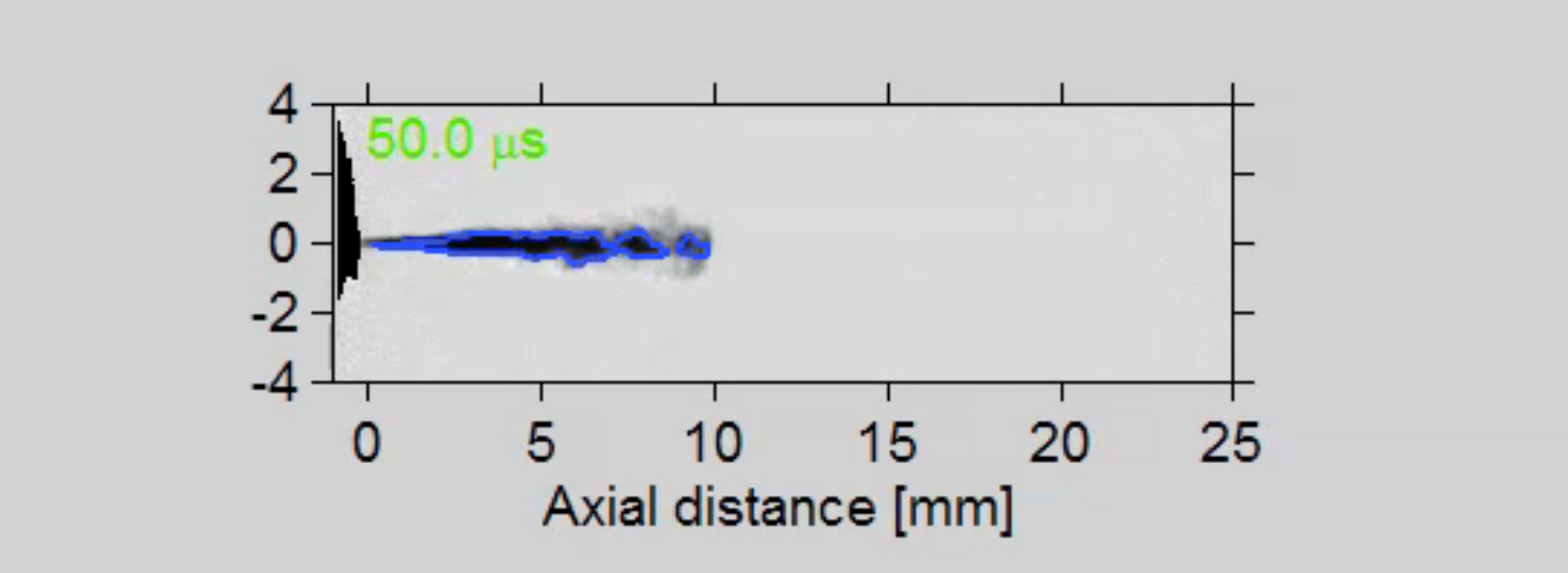}
    }
    \subfigure{
        \includegraphics[width=0.45\textwidth,trim={105 75 10 540},clip=,frame]{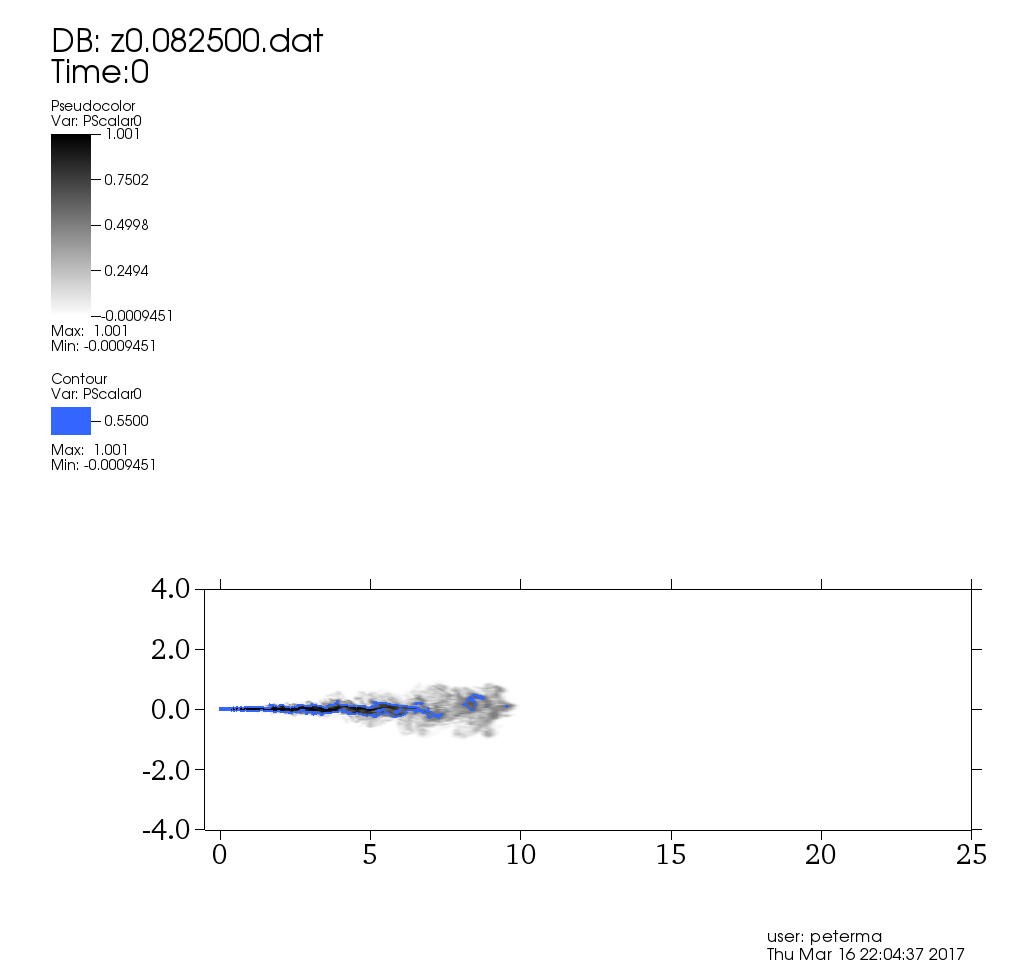}
    }
    \subfigure{
        \includegraphics[width=0.45\textwidth,trim={75 32 85 15},clip=]{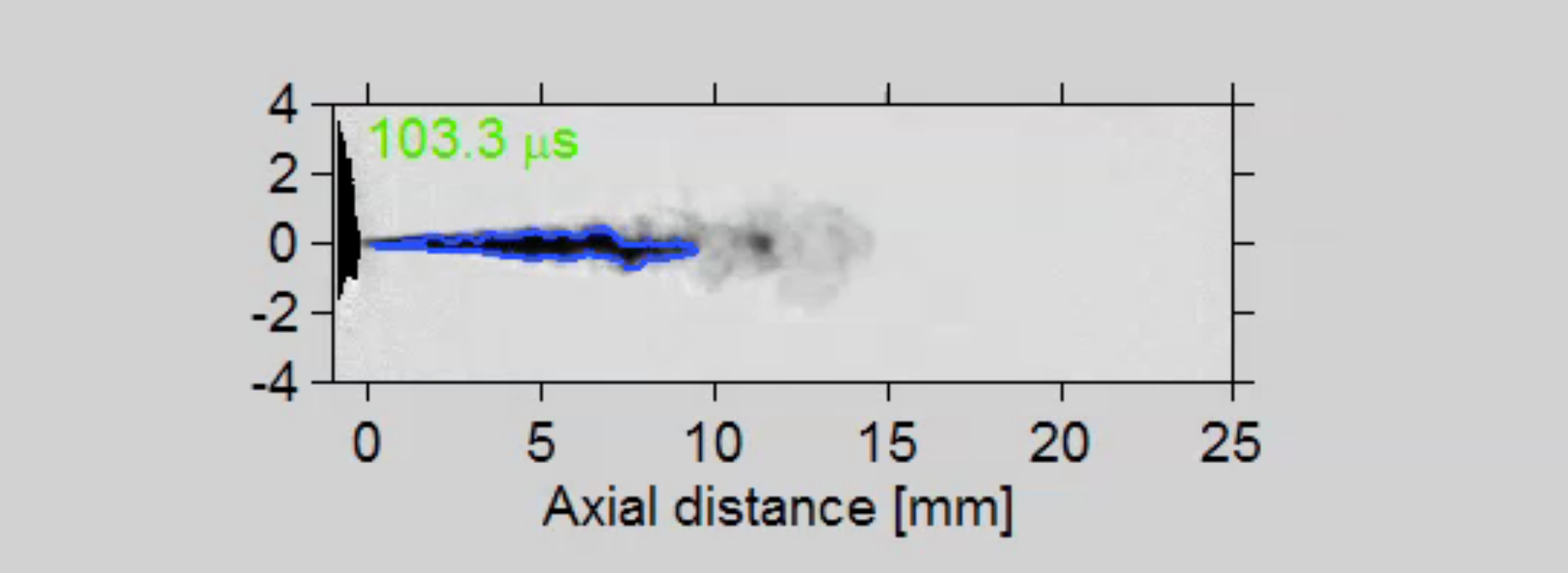}
    }
    \subfigure{
        \includegraphics[width=0.45\textwidth,trim={105 75 10 540},clip=,frame]{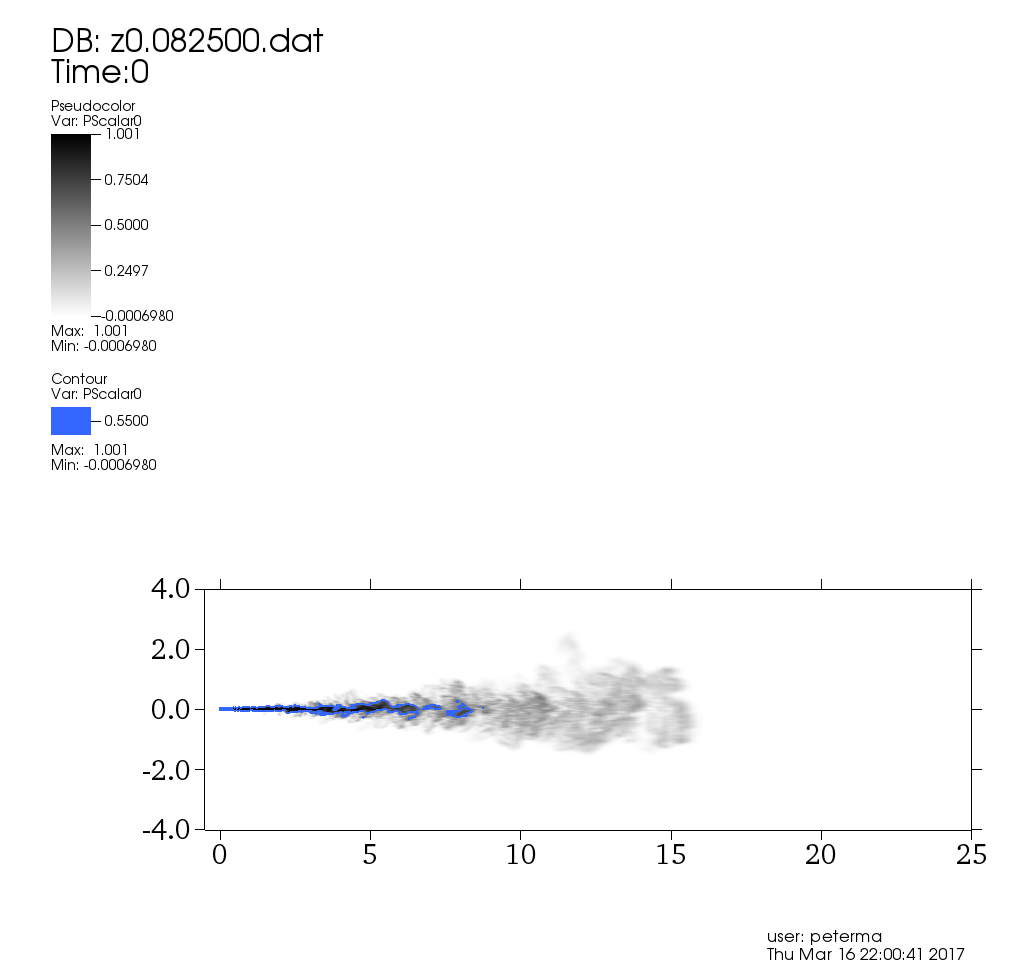}
    }
    \subfigure{
        \includegraphics[width=0.45\textwidth,trim={75 32 85 15},clip=]{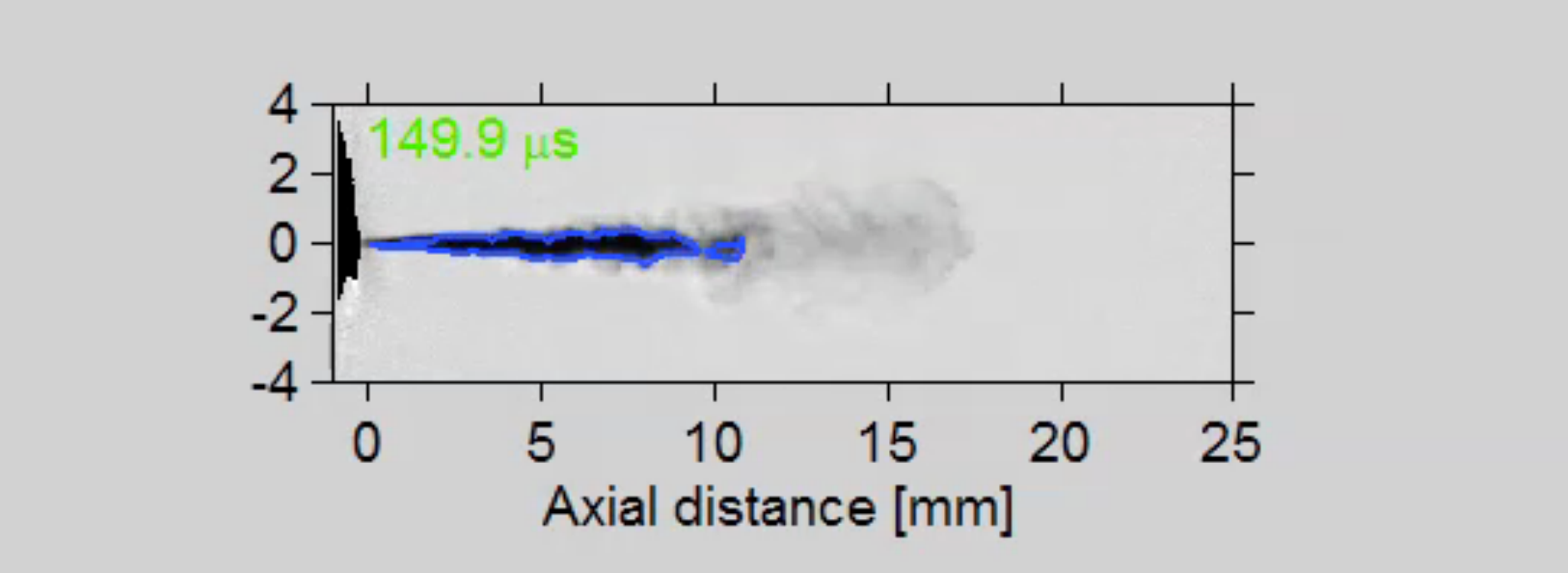}
    }
    \subfigure{
        \includegraphics[width=0.45\textwidth,trim={105 75 10 540},clip=,frame]{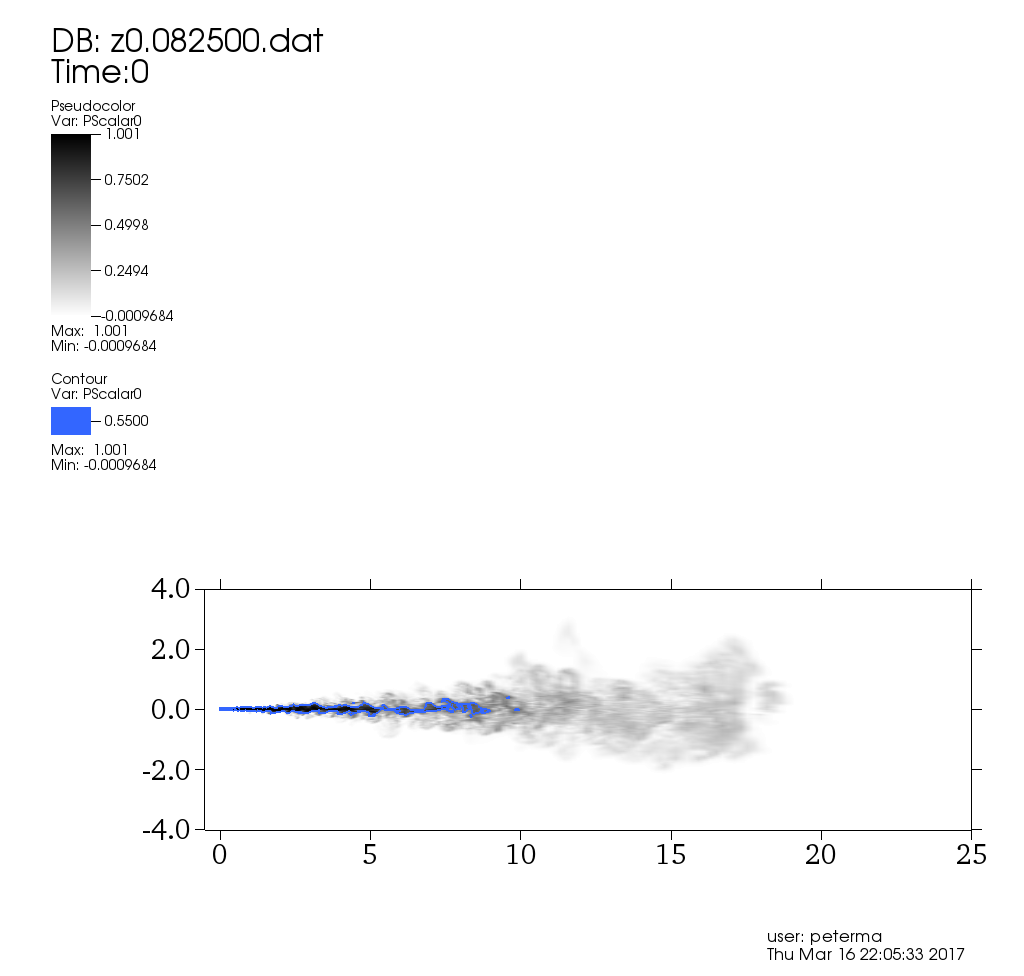}
    }
    \subfigure{
        \includegraphics[width=0.45\textwidth,trim={75 32 85 15},clip=]{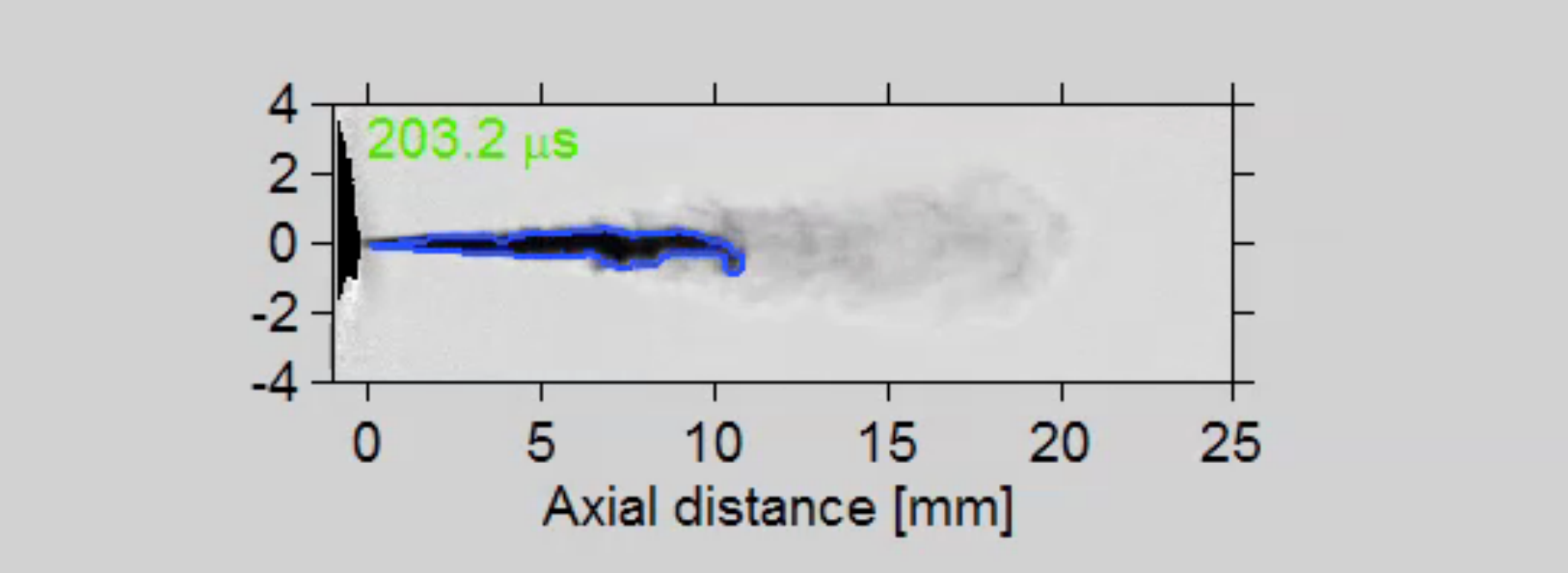}
    }
    \subfigure{
        \includegraphics[width=0.45\textwidth,trim={105 75 10 540},clip=,frame]{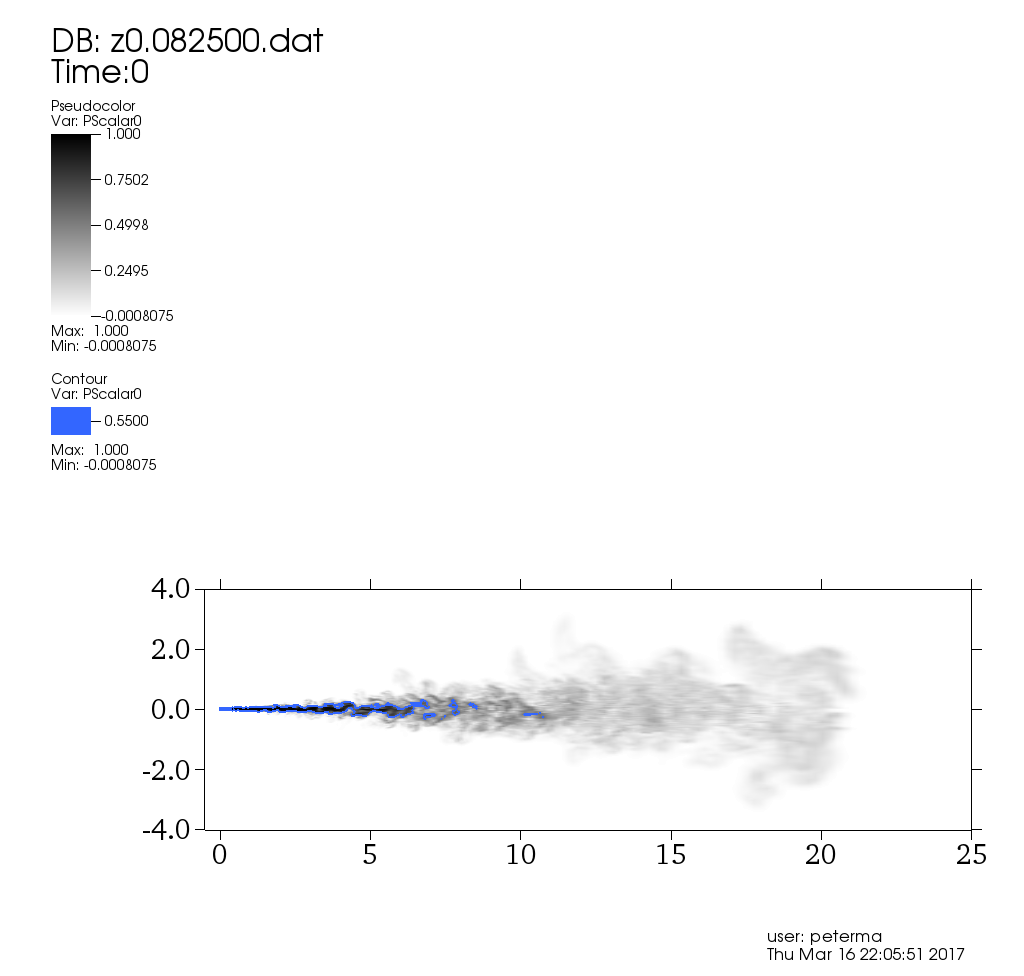}
    }
    \setcounter{subfigure}{0}
    \subfigure[Experiment]{
        \includegraphics[width=0.45\textwidth,trim={75 32 85 15},clip=]{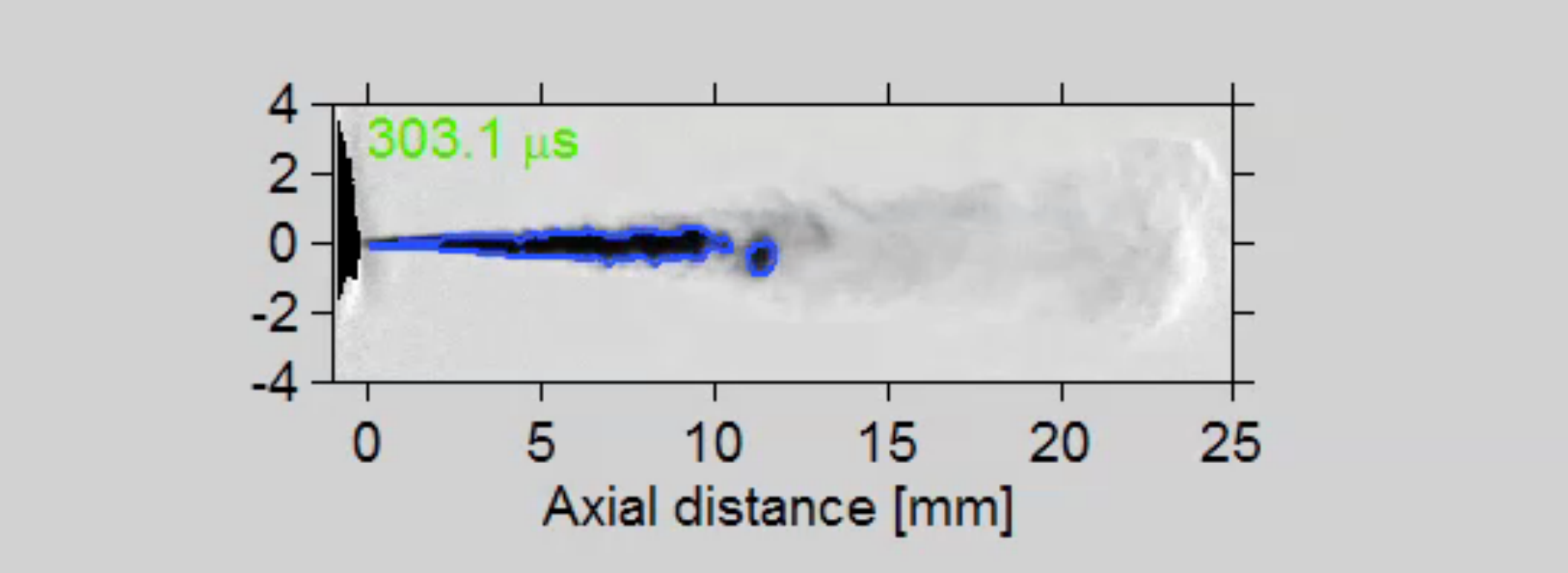}
    }
    \subfigure[LES]{
        \includegraphics[width=0.45\textwidth,trim={105 75 10 540},clip=,frame]{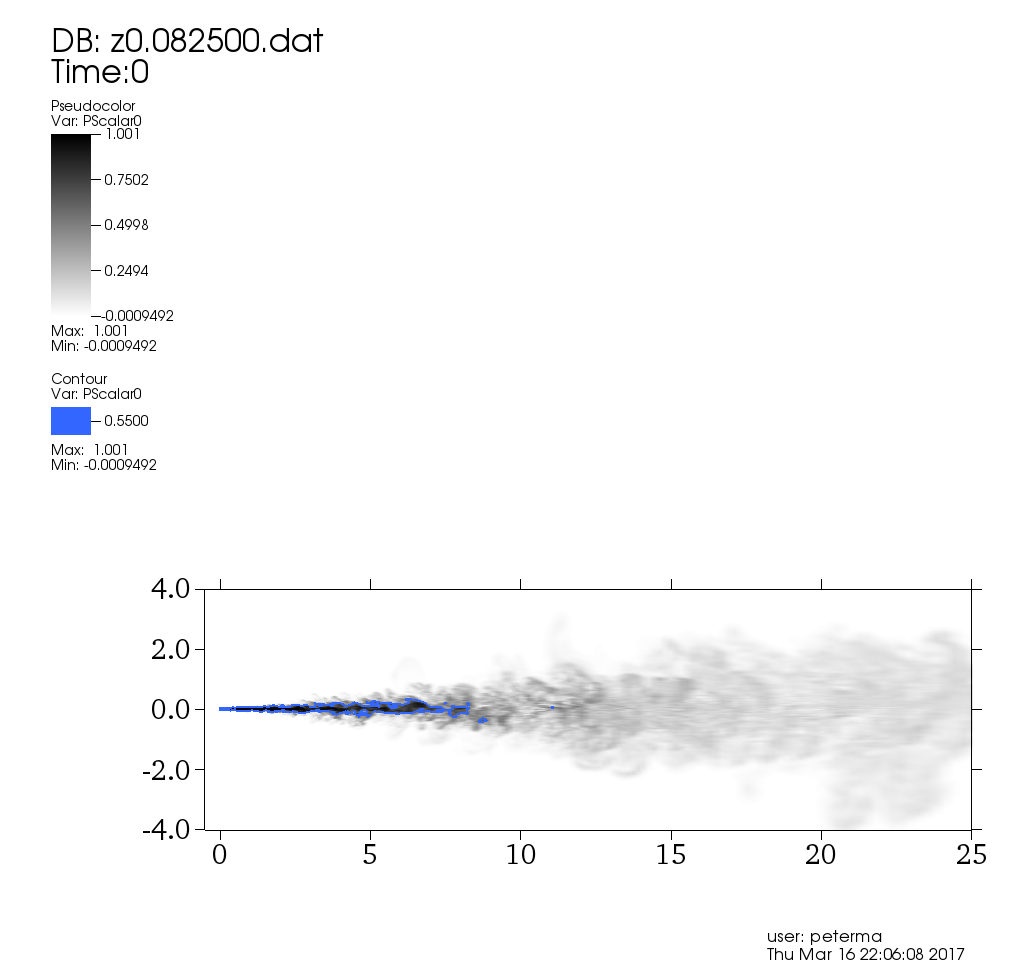}
    }
    \caption{Comparison between (a) experiment and (b) LES for the injection sequence. Experimental images are obtained using diffused back-illumination method with grayscale intensity threshold to indicate liquid region~\cite{pickett2011measurement, manin2012evaluation}. Fuel mass fraction contours at the center-plane are shown in LES results with dense fluid region indicated by mass fraction with a value of 0.6. Units for spatial dimensions are in mm.\label{fig:comparison}}
\end{figure*}

\Cref{fig:comparison} shows a temporal sequence of the injection process from the experiment and current numerical simulation. From this figure it can be seen that the dense fuel jet remains stable in the vicinity of the injector nozzle, for $x < 3$~mm. The jet starts to break up downstream of this region and instabilities at the shear layer can be clearly seen. Further downstream, the fuel mixes with the ambient gas and large eddies are present. The spreading angle of the injected fuel is narrow in the injector near-region and becomes wider further downstream where turbulent mixing is present. This is similar to numerical results reported in other studies~\cite{lacaze2015analysis,knudsen2016compressible,matheis2016multi}.

The simulation results are qualitatively in good agreement with the experimental images taken by the diffused back-illumination (DBI) method~\cite{pickett2011measurement, manin2012evaluation} in terms of fuel penetration and overall spreading rate. However, as can be seen from \cref{fig:comparison}, the fuel jet behaves differently in the near injector region. The dense liquid regions are indicated by the blue curves for both the experiment and LES in \cref{fig:comparison}. The liquid region in the experiment is approximately determined by the darkness of the grayscale intensity~\cite{manin2012evaluation}. A value of 0.6 for the fuel mass fraction is used for the determination of the liquid region in the numerical simulation. It can be clearly seen that the experimental results have a wider spreading angle near the injector. Several possible reasons can explain this discrepancy between experiments and simulation results. One is that the flow inside the injector pipe is not considered in the current simulation. By considering a  turbulent inflow profile, stronger shear layer instabilities are expected. Another reason could be due to the limitation of the diffused interface method so that the flow physics in the near injector region is not modeled correctly since interfaces and droplets are completely excluded. However, there is no direct way of determining the liquid region with a diffused interface method. Note that a methodology is developed by Manin~et~al.~\cite{manin2012evaluation} to use DBI images to experimentally determine the liquid penetration length, and this methodology is shown to give longer penetration compared to Mie scattering. 

\begin{figure}[!h!]
    \centering
    \includegraphics[width=0.95\columnwidth]{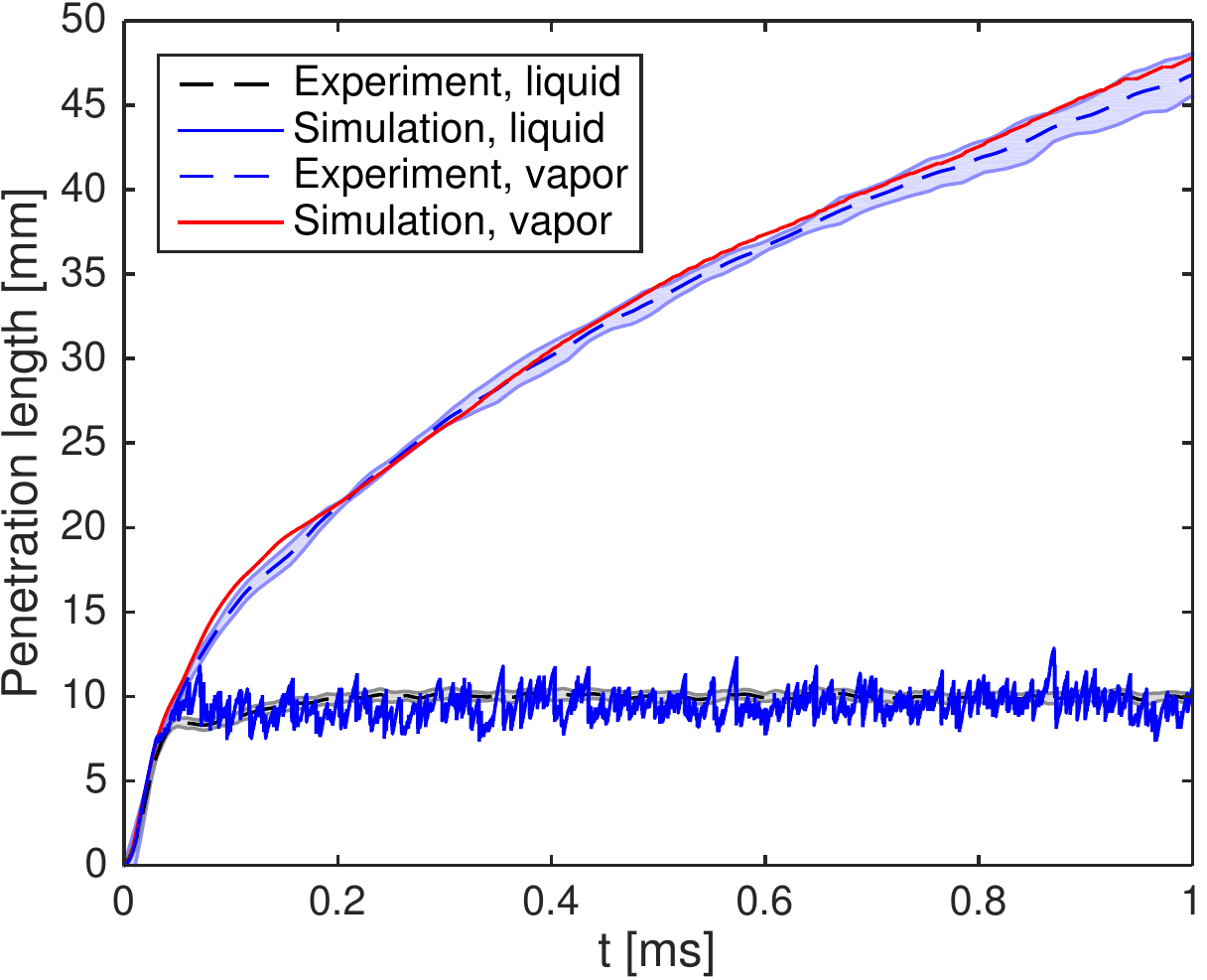}
    \caption{Liquid and vapor penetration lengths predicted by LES in comparison with experimental data~\cite{pickett2009visualization}. Experimental liquid penetration length was determined from high-speed Mie scattering using a 3\% threshold of maximum intensity. The experimental vapor penetration length was determined from Schlieren imaging. Thresholds of 0.6 and 0.01 for the fuel mass fraction were used for the determination of liquid and vapor penetration lengths, respectively, in LES.\label{fig:pl}} 
\end{figure}

The liquid and vapor penetration lengths are extracted from the simulation results using a threshold value of 0.6 and 0.01 for the fuel mass fraction, respectively. The results up to 1~ms after injection are shown in \cref{fig:pl}. The experimental vapor penetration length determined from Schlieren imaging and liquid penetration length from Mie scattering~\cite{pickett2011engine,pickett2009visualization} are also shown for comparison. It can be seen that for the vapor penetration an excellent agreement with measurements is obtained. It was found that the utilization of the inflow boundary condition with a time-dependent fuel mass flux is essential for the accurate prediction. Another simulation with a constant mass flux without the initial ramp-up yielded longer penetration length which is consistent with other studies~\cite{pickett2011engine}. The sensitivity of the criterion for the determination of the vapor penetration was studied, confirming that the so computed vapor penetration length was insensitive to threshold values between 0.005 and 0.1.

For the validation of the liquid penetration length, a threshold value of 0.6 is used for the fuel mass fraction, providing excellent agreement with the measurements (see~\cref{fig:pl}). However, this threshold value is somewhat arbitrary. The sensitivity of the liquid threshold value was also tested, and it was found that for mass fraction values from 0.95 to 0.4 the predicted length varies between 5~mm to 15~mm. It was pointed out in the experimental investigations~\cite{pickett2011measurement,manin2012evaluation} that the experimentally measured liquid penetration length is sensitive to the choice of the measurement method, the optical setup within one method, the criteria of liquid determination, and the actual geometry of the injector. Indeed, the resolution in either the simulation or the experiment is still not close to fully resolve the interfacial flows near the injector nozzle, if multi-phase flows do exist under these conditions.

\begin{figure}[!htb!]
    \centering
    \includegraphics[width=0.95\columnwidth]{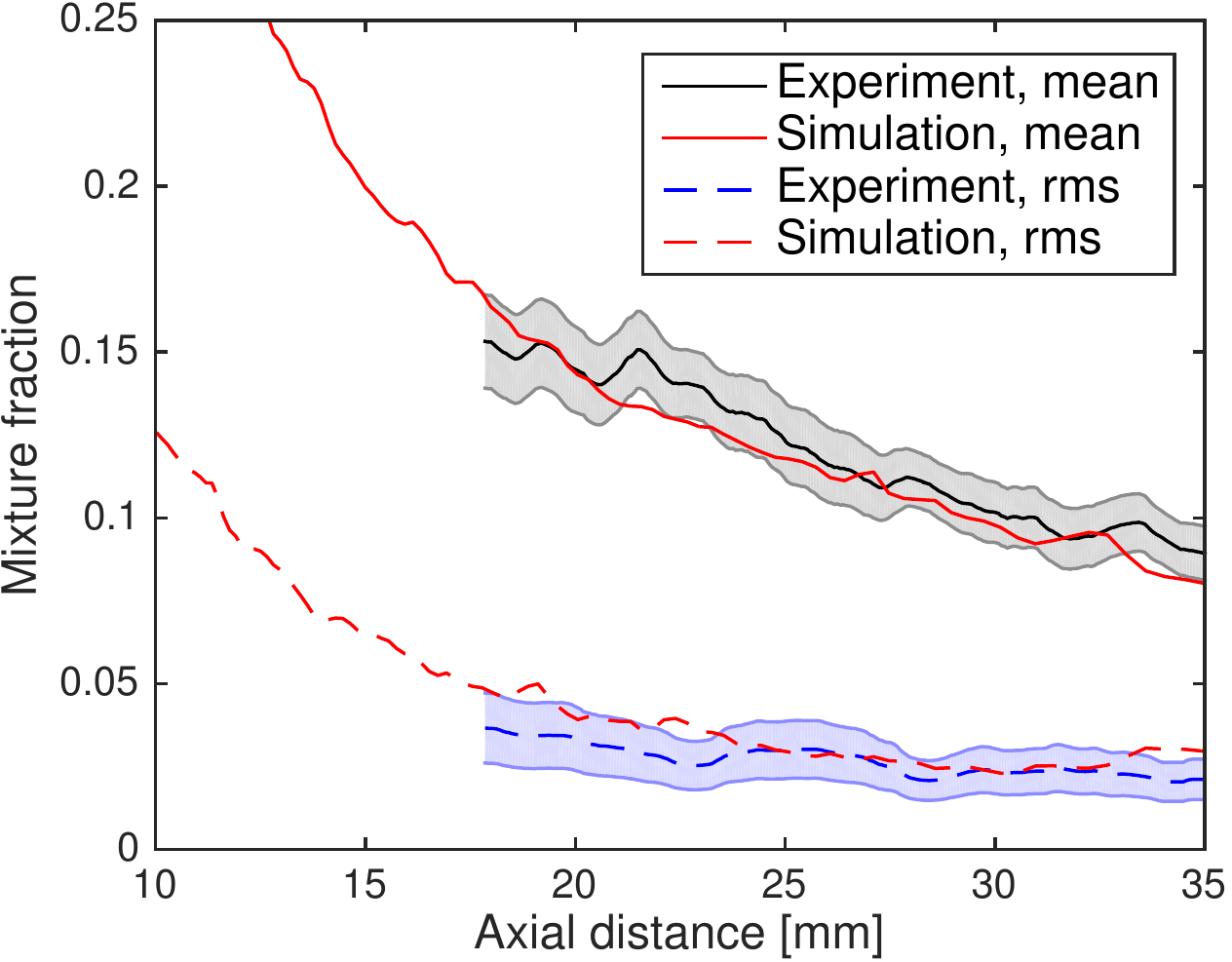}
    \caption{Axial profiles of mean and rms values of mixture fraction at the center-line in comparison with the experimental data measured by Rayleigh scattering~\cite{pickett2011relationship}.\label{fig:axial}} 
\end{figure}

\begin{figure}[!h!]
    \centering
    \subfigure{
        \includegraphics[width=0.95\columnwidth,clip=]{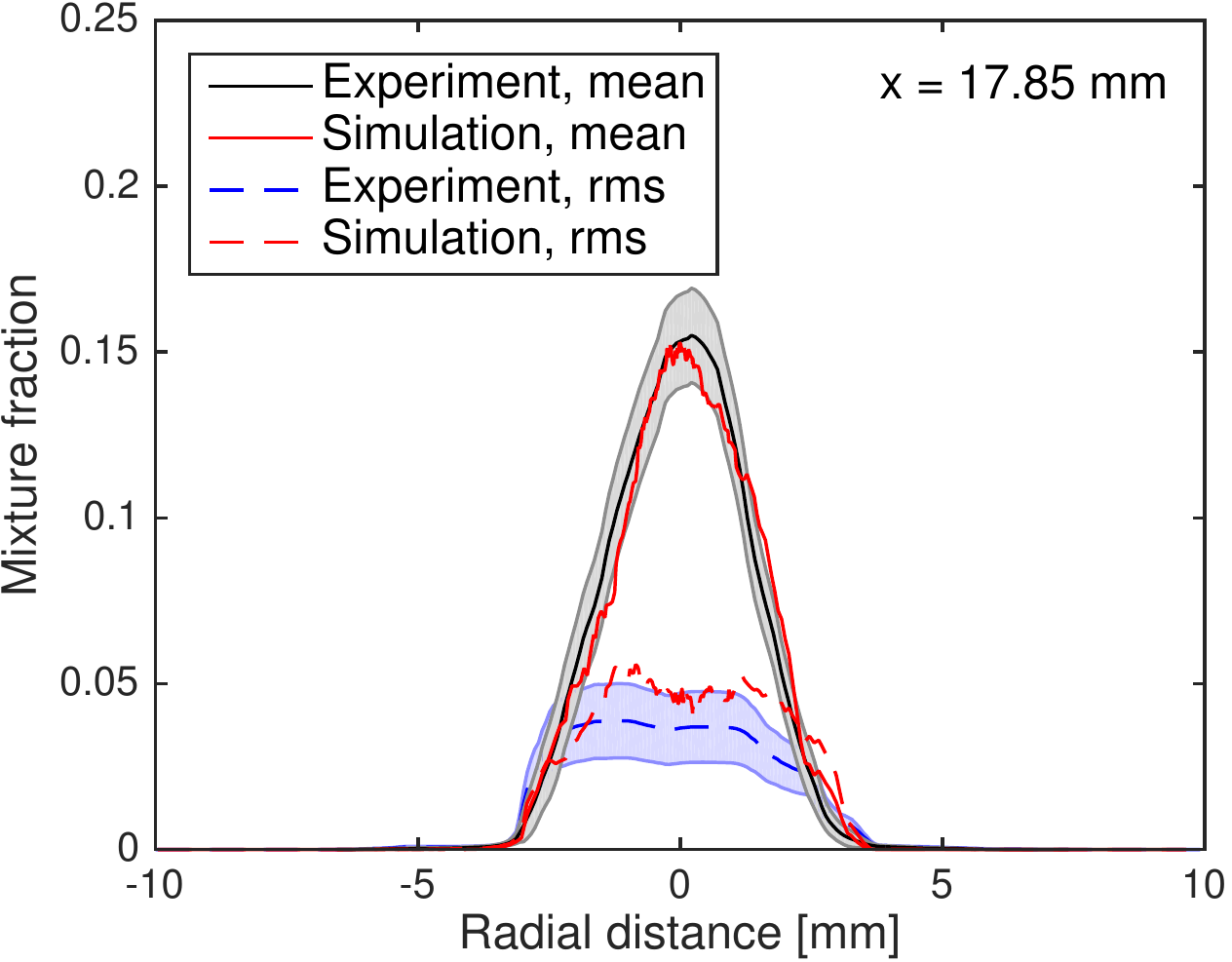}
    }
    \subfigure{
        \includegraphics[width=0.95\columnwidth,clip=]{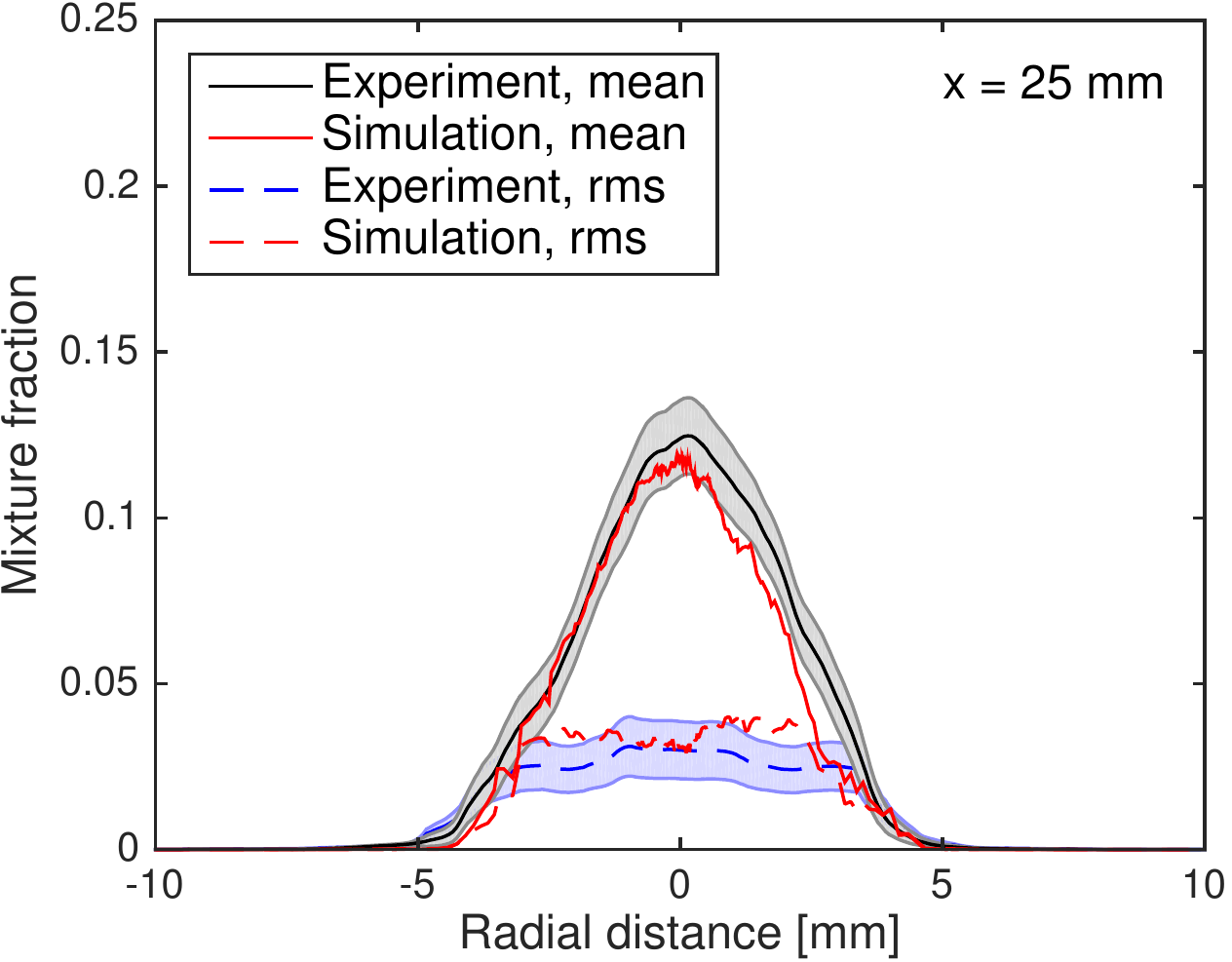}
    }
    \subfigure{
        \includegraphics[width=0.95\columnwidth,clip=]{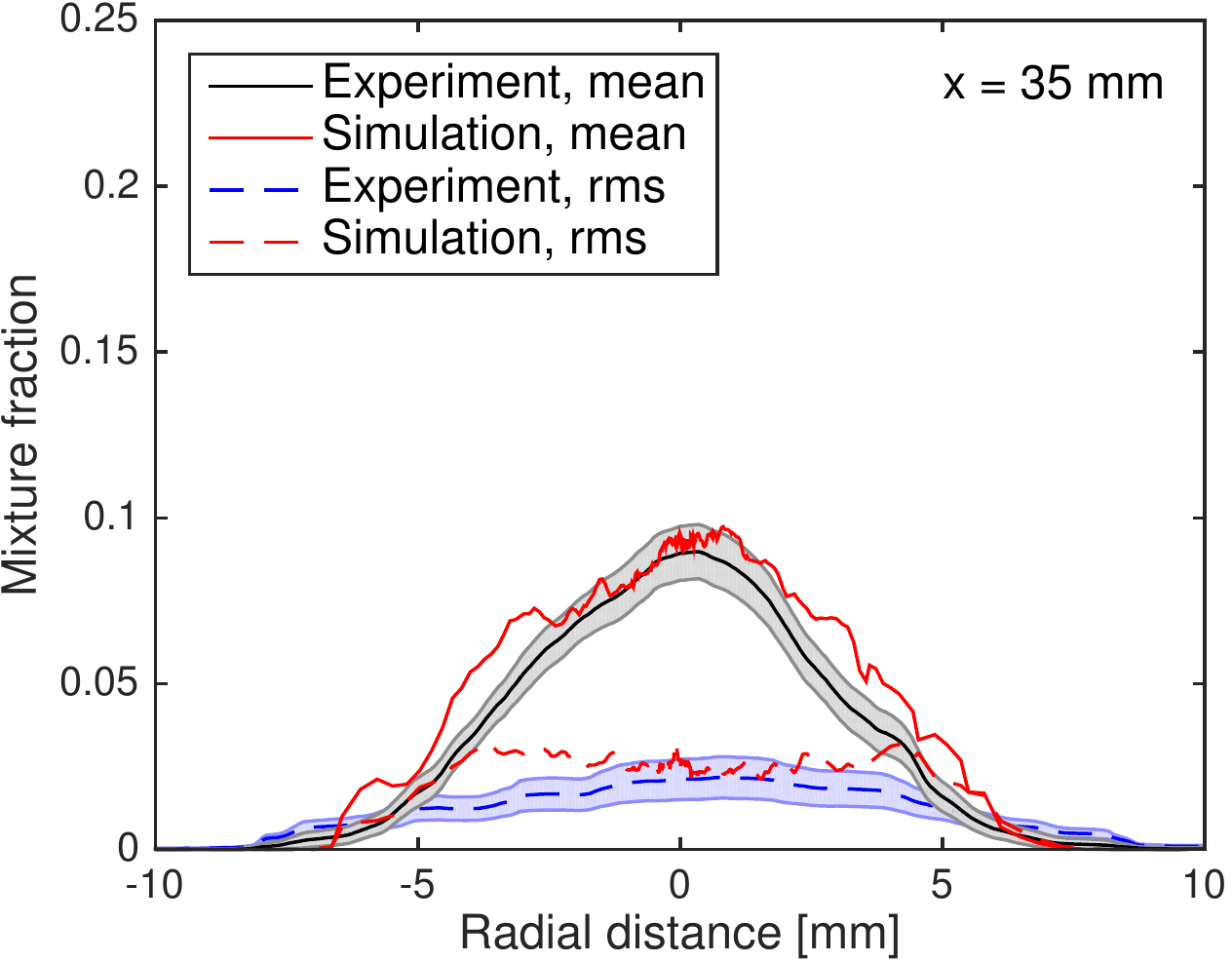}
    }
    \caption{Radial profiles of mean and rms values of mixture fraction at three different axial locations in comparison with the experimental data measured by Rayleigh scattering~\cite{pickett2011relationship}.\label{fig:radial}}
\end{figure}

The flow structures and mixing behaviors of the injection process further downstream are compared to the measurements of mixture fraction by Rayleigh scattering~\cite{pickett2011relationship}. The results are shown in \cref{fig:axial} and \cref{fig:radial}. Multiple injections in the experiments provide ensemble-averaged statistics. In the simulation, the statistics of the steady period of injection are obtained by temporally averaging between 0.6~ms and 1.2~ms after the injection. \Cref{fig:axial} shows the comparison of mean and root-mean-square (rms) values of mixture fraction at the center-line of the domain. Good agreement is obtained for both the mean and rms values and the simulation results fall within experimental uncertainties. \Cref{fig:radial} shows a comparison of the radial mixture fraction distribution at three different axial locations ($x = 17.85$, 25, and 35~mm). As can be seen in \cref{fig:radial}, there is a good agreement in the mean values of the mixture fraction at all three locations, while the simulation predicts slightly higher rms values compared to the experimental data. These results along with the excellent agreement of the vapor penetration length as presented in~\cref{fig:pl}, show that the current numerical method is capable of predicting the turbulent mixing process between fuel and surrounding environment downstream of the injector after the dense liquid fuel is fully disintegrated.

\section{Summary and Conclusions}\label{sec:conclusions}
The need for improved engine efficiencies has motivated the development of high-pressure combustion systems, in which operating conditions achieve and exceed critical conditions. Associated with the transcritical conditions are large thermodynamic gradients as the fluid undergoes mixing and possibly phase transitions. Accurately simulating these real-fluid environments remains a challenge. In this study, a diffused interface method is presented for the modeling of the fuel injection process under conditions relevant to diesel engines. Compressible multi-species conservation equations are solved in conjunction with the Peng-Robinson state equation and real-fluid transport properties. A finite-volume method with an entropy-stable scheme is utilized to robustly and accurately simulate real-fluid flows with strong nonlinearities. A recently developed double-flux model is employed to eliminate spurious pressure oscillations associated with transcritical flows.

LES-calculations are performed to simulate the ECN Spray A target case~\cite{pickett2011engine}. Simulation results are analyzed and compared to experimental measurements. The simulation results are qualitatively in good agreement with the experimental images. Different flow dynamics are found in the vicinity of the nozzle where the simulation predicts a narrower spreading angle of the fuel jet. Excellent agreement in vapor penetration length is obtained. The liquid penetration length is found to be sensitive to the selected criterion. Comparisons of the mixture fraction in the turbulent mixing portion of the jet show that the developed modeling capability is able to accurately predict the mixing behavior after the dense liquid jet fully disintegrates.

Observed differences in the flow-field behavior near the injector requires further investigations both numerically and experimentally. Simulations utilizing sharp interface method may provide insights into this question, which is the focus of ongoing work.

\section*{Acknowledgments}
Financial support from the U. S. Army Research Laboratory through the Cooperative Agreement W911NF-16-2-0170 is gratefully acknowledged.


 \end{document}